# Signatures of exciton condensation in a transition metal dichalcogenide

Anshul Kogar[1], Melinda S. Rak[1], Sean Vig[1], Ali A. Husain[1], Felix Flicker[2], Young Il Joe[1], Luc Venema[1], Greg J. MacDougall[1], Tai C. Chiang[1], Eduardo Fradkin[1], Jasper van Wezel[3], Peter Abbamonte[1]

[1] Department of Physics and Seitz Materials Research Laboratory, University of Illinois, Urbana, IL, 61801, USA
[2] Department of Physics, University of California, Berkeley, California 94720, USA
[3] Institute of Physics, University of Amsterdam, 1098 XH Amsterdam, The Netherlands

**Bose condensation has shaped our understanding of macroscopic quantum phenomena, having been realized in superconductors, atomic gases, and liquid helium. Excitons are bosons that have been predicted to condense into either a superfluid or an insulating electronic crystal. Using the recently developed momentum-resolved electron energy-loss spectroscopy (M-EELS), we study electronic collective modes in the transition metal dichalcogenide semimetal, 1$T$-TiSe$_2$. Near the phase transition temperature, $T_C$=190K, the energy of the electronic mode falls to zero at nonzero momentum, indicating dynamical slowing down of plasma fluctuations and crystallization of the valence electrons into an exciton condensate. Our study provides compelling evidence for exciton condensation in a three-dimensional solid and establishes M-EELS as a versatile technique sensitive to valence band excitations in quantum materials.**

If a fluid of bosons is cooled to sufficiently low temperature, a significant fraction will condense into the lowest quantum state, forming a Bose condensate. Bose condensation is a consequence of the even symmetry of the many-body wave function of bosons under particle interchange, and allows for the manifestation of macroscopic quantum phenomena, the most striking being superfluidity.

Traditionally, Bose condensates are said to come in two types. Bose-Einstein condensates (BECs) occur in systems of stable bosons, such as dilute atomic gases or liquid $^4$He, while in Bardeen-Cooper-Schrieffer (BCS) condensates the bosons are formed of bound states of fermions, such as the Cooper pairs in a superconductor or superfluid $^3$He. In reality, all such bosons (e.g., a $^4$He atom) are composed of bound fermions, and BEC and BCS are just different limits of the same phase of matter (*1*). Experiments on such condensates have shaped the modern understanding of quantum field theory (*1*).

Excitons are bosons that are bound states between an electron and hole in a solid, and were predicted long ago to Bose condense (*2,3,4*). Because of their light mass and high binding energy, exciton condensates should be stable at higher temperature than traditional BEC or BCS phases (*5,6*). Different theories predict that a Bose condensate of excitons could be a superfluid (*5*) or innately insulating (*7*), so there is tremendous need for experimental input. Identifying an exciton condensate in nature could have a profound impact on future understanding of macroscopic quantum phenomena, as well the classic problem of the metal-insulator transition in band solids, in which exciton condensation has long been believed to play a fundamental role (*2,3,4*).

Condensed phases of photogenerated excitons have been realized in semiconductor quantum wells in resonance with a Fabry-Perot cavity which, although not fully thermally equilibrated, have exhibited evidence for transient superfluidity (*8*). Excitonic phases have also been realized in quantum Hall bilayers in a perpendicular magnetic field (*9*). Although the order in these two-dimensional structures is not strictly long-ranged, and the order parameter cannot be measured directly, compelling evidence for excitonic correlations has been observed in Coulomb drag experiments (*9*). Despite these achievements, there is a great need to identify an exciton condensate in a fully thermalized, three-dimensional system in which the order is long-ranged.

An ideal approach would be to identify a material in which an exciton condensate forms "naturally." Long ago, a BCS condensate of excitons was predicted to arise spontaneously in semimetals in which an indirect band gap is tuned close to zero (Fig. 1) (*2,3,4*). This condensate is expected to break a spatial symmetry, rather than the *U*(1) symmetry broken by a superconductor, and in the absence of pinning should exhibit perfect conductivity without a Meissner effect (10). This phase can be thought of as a solid crystal of excitons, which early authors dubbed "excitonium" (4), and is the two-band analogue of the Wigner crystal instability of an interacting electron gas (10). This condensate is closely related to that in bilayer quantum wells (9), the coherence developing between electrons and holes in different bands (Fig. 1) rather than in different layers. If found, this exciton condensate would be three-dimensional, guaranteed to reside in thermodynamic equilibrium, and could potentially be stable close to room temperature, facilitating fundamental studies of this many-body phenomenon (*2,3,4,7*).

The challenge is that an exciton condensate in a solid is nearly impossible to distinguish from a Peierls charge density wave. A Peierls phase is a spontaneous crystal distortion driven by the electron-phonon interaction (*11*), and is unrelated to exciton formation. But the two phases have the same symmetry and similar physical observables, such as the existence of a superlattice and the opening of a single-particle energy gap. Although many compelling candidate excitonic materials have been identified (*12,13,14,15,16,17*), there has not been an experimental means to distinguish between Peierls and excitonic ground states, i.e., to determine whether an electron-hole condensate is present.

Here, we demonstrate the existence of an exciton condensate in 1*T*-TiSe$_2$ by measuring its collective excitations using momentum-resolved electron energy-loss spectroscopy (M-EELS) (*18*). The defining characteristic of a Peierls phase is the presence of a soft phonon whose energy falls to zero at finite momentum, $q_0$, at its phase transition temperature, $T_C$ (*11,19*). Below $T_C$, this phonon splits into amplitude and phase branches, the latter (if it remains gapless) serving as the Goldstone mode (*11,19*).

In contrast, the Goldstone mode of an excitonic condensate corresponds to translation of the electronic crystal with respect to the atomic lattice and was dubbed "excitonic sound" by early authors (*20*). The precursor to this mode at $T > T_C$ is an electronic, plasmon-like excitation, which defines the excitonic transition by falling to zero energy at $T_C$. This electronic mode can mix with phonons at low energy but should disperse to the plasma frequency as $q \to 0$ (*20*). In other words, a material could be established to contain an exciton condensate if it exhibited a "soft plasmon," analogous to the soft phonon of a Peierls transition (*4,17,20*).

The observation of an electronic mode that falls to zero energy at finite momentum would indicate that the energy to create an electron-hole pair is zero, which is unambiguous evidence for condensation of electron-hole pairs, i.e., the development of a macroscopic population of excitons in the ground state. In the following, we will refer to this electronic mode as a plasmon, although elucidating its exact character, including whether it is longitudinal, transverse, or a mixture of the two, requires further theoretical and experimental study.

One of the most promising candidate excitonic materials is the transition metal dichalcogenide (TMD) semimetal, 1*T*-TiSe$_2$, whose vanishing indirect band gap is optimal for realizing an exciton condensate (*2,3,4,17*). At $T_C$ = 190 K, TiSe$_2$ exhibits a resistive anomaly and forms a $2a \times 2a \times 2c$ superlattice whose wave vector connects the Se 4*p* valence band at the $\Gamma$ point to the Ti 3*d* conduction band at the *L* point, leading many authors to identify the material as residing in an excitonic state (*17,21,22*). However, TiSe$_2$ also exhibits a sizeable lattice distortion (*17,23*), leading others to argue that it is a phonon-driven Peierls-like phase (*24,25,26,27,28*).

M-EELS studies of TiSe$_2$ were carried out at 50 eV beam energy using an Ibach-type HR-EELS spectrometer (*29*). Momentum resolution was achieved by precisely aligning the spectrometer to the rotation axes of a low-temperature sample goniometer operated with a custom control system adapted from x-ray diffraction (*18*). M-EELS measures the charge dynamic structure factor of a surface, $S(\mathbf{q},\omega)$, which is proportional to the imaginary part of the frequency- and momentum-dependent charge susceptibility, $\chi''(\mathbf{q},\omega)$ (18). This quantity is proportional to the dielectric loss function $-\text{Im}[1/\varepsilon(\mathbf{q},\omega)]$ (*30*), so M-EELS can be thought of as momentum-resolved infrared (IR) spectroscopy. M-EELS is more sensitive to valence band excitations than inelastic x-ray or neutron scattering, which mainly couple to lattice modes, and is particularly suited to studies of electronic collective excitations (*18*). The results presented here were reproduced five times on different cleaved crystals, each of which provided reproducible data for about 40 hours under ultra-high vacuum conditions (Fig. S6) (*35*).

M-EELS measurements of TiSe$_2$ are consistent with experimental results obtained with other methods. Figures 2, A and B show static (ω=0) M-EELS maps of momentum space taken with an energy resolution of 6 meV. Superlattice reflections, indicating breaking of translational symmetry and the development of a static order parameter, appear below $T_C$ at wave vectors $q$ = (0.5, 0) and (0.5, 0.5) (note that, because M-EELS is a surface probe, momenta are indexed using only the two components parallel to the surface). The peak locations are in quantitative agreement with x-ray, neutron, and electron diffraction studies (*17,23*)

Frequency-dependent M-EELS spectra, taken at a fixed momentum $q$=0 for a series of temperatures, are shown in Fig. 2C. At $T$ = 300 K, a highly damped electronic mode is observed at 82 meV whose energy and linewidth decrease when cooling through $T_C$. The mode energy increases again at low temperature, reaching 47 meV at $T$ = 17 K. These changes agree quantitatively with previous IR spectroscopy studies (*31,32,33*), which identified this excitation as a free carrier plasmon, because its energy is significantly higher than the highest optical phonon at 36 meV (*27,34*). These changes in plasmon energy and linewidth were interpreted as a decrease in carrier density and Landau damping caused by opening an energy gap at $T_C$ (*31,32,33*).

The plasmon in Fig. 2C is the fundamental electronic collective mode of TiSe$_2$. The question of whether an exciton condensate is present reduces to asking whether this excitation exhibits the behavior of a soft mode at $T_C$.

At $T > T_C$ (Fig. 3A), the momentum dependence of the plasmon in TiSe$_2$ is unremarkable. Its energy and linewidth increase along the (1,0) direction with increasing $q$, consistent with the usual Lindhard description of a conductor (*30*). Upon cooling to $T$ = 185K ~ $T_C$ (Fig. 3B), the plasmon becomes anomalous: its energy decreases with increasing $q$ until it merges with the zero-loss line for $q >$ (0.3,0). Near the ordering wave vector, $q_0$ = (0.5, 0), the excitation is gapless within the resolution of the measurement, the M-EELS spectrum exhibiting a power law form, $S \sim \omega^{-1}$ (*35*). At momenta $q >$ (0.7, 0), the electronic mode emerges again, increasing in energy to its maximum value at $q$ = (1,0). This behavior is exactly what is expected of a soft mode (*11,19*).

Cooling further to $T$ = 100 K (Fig. 3C), well below $T_C$, the excitation hardens, forming a dispersing mode that is resolvable at $q_0$ = (0.5, 0) as a shoulder on the elastic line. By $T$ = 17 K (Fig. 3D), the mode hardens to 50 meV and its energy and linewidth become approximately momentum-independent.

This unusual plasmon mode was not observed in inelastic x-ray (*36*) and neutron (*34*) scattering experiments, which lack the valence sensitivity of M-EELS (*18*). We quantified its behavior by fitting the spectra with a series of Gaussians and Lorentzian functions for the elastic peak, several phonons, and the plasmon mode, including both Stokes and anti-Stokes (energy gain) features (Fig. 4) (*35*).

The behavior displayed in Fig. 4 is that of a soft mode, demonstrating that electron-hole pairs in TiSe$_2$ are condensing at $T_C$, forming a Bose condensate of excitons. The overall picture that emerges is that, at $T > T_C$, TiSe$_2$ is an ordinary semimetal with a free carrier plasmon at $\omega_p$ = 82 meV. As the system is cooled toward $T_C$ the plasma fluctuations slow down, particularly at $q = q_0$. The electron spectral function as measured by angle-resolved photoemission (ARPES) begins to show the precursor to an energy gap (*17,22,25,38*), and the plasmon shifts to lower frequency and becomes overdamped (Figs. 3, 4). Slightly above $T_C$, the correlation function, $S(q_0,\omega)$, exhibits power laws indicating dynamical critical fluctuations (35). At $T = T_C$ a single-particle gap opens, the mean plasmon energy reaches zero and a finite population of excitons forms, leading to a static ($\omega = 0$), resolution-limited order parameter reflection at $q_0$ (Fig. 2B). Below $T_C$ the population of the condensate becomes macroscopic, a spectral gap in ARPES is established, and the plasmon hardens into an amplitude mode that is only weakly dispersive (Fig. 4). This excitation, identified as a plasmon in IR studies (31,32,33), is better thought of as an exciton, because it represents a quantized modulation of the amplitude of the condensate. It is an open question to what extent this excitation exhibits the character of a conventional plasmon as $q \to 0$. The associated phase mode was not observed in M-EELS and may not exist as a distinct excitation in a condensate with period $2a$ because its order parameter is real. A microscopic model is needed to know for certain.

The lattice degrees of freedom are not inert in TiSe$_2$, which exhibits a sizeable lattice distortion (17,23), implying the existence of a soft phonon. As illustrated in Fig. 4, only one of the nine phonon branches in TiSe$_2$ participates in the phase transition—the transverse $L_2$ mode, which goes soft at $T_C$ (34,36). The $L_2$ distortion may be a passive consequence of the excitonic state. However, several authors have suggested that, if an exciton condensate is present, this transverse lattice distortion may help stabilize it (37,38). In this situation, the $L_2$ phonon and soft plasmon must interact.

Although experiments lack the resolution to say for certain, we speculate that the plasmon and $L_2$ phonon interact as sketched in Fig. 1C. At $q = 0$, the two excitations should be distinct. If both contribute to the transition, at $T = T_C$ both will soften toward zero energy at $q = q_0$. The electron-phonon interaction will cause the two modes to mix, leading to an avoided crossing near $q_0$ where the hybrid excitations have equal phonon and plasmon character. Only one will act as a soft mode, the other remaining at finite energy. Note that this mixing should only be nonzero if the electronic mode itself also exhibits some transverse character, though additional studies are required to ascertain this. The condensed electron-hole pairs are then dressed by phonons, creating a lattice distortion that coexists with the exciton condensate.

Nevertheless, TiSe$_2$ is qualitatively different from Peierls materials, which behave as illustrated in Fig. 1D. The only soft mode in a Peierls transition is a phonon, the plasmon exhibiting no change near $T_C$. Examples of Peierls materials are NbSe$_2$ and TaS$_2$ in which strong metallic screening prevents excitons from forming, and the plasmon resides at high energy, $\omega_p \sim 1\,\text{eV}$, and is unaffected by the phase transition (*39*). Weak mixing between plasmon and phonon modes still occurs because the electron-phonon interaction is nonzero. But the degree of mixing is of order $g/\omega_p$, where $g$ is a plasmon-phonon coupling constant. Using the theory described in Ref. (*40*), we estimate $g \sim 6$ meV in NbSe$_2$, implying $g/\omega_p \sim 0.006$, meaning the soft mode is less than 1% electronic in character, compared to ~50% for the case of an excitonic state. Hence, the distinction between an exciton condensate and a Peierls phase is, ultimately, a quantitative matter. TiSe$_2$ is special in that the electronic character of its ordered state is orders of magnitude larger than any other material, setting it apart as hosting a crystalline state of solid excitonic matter.

**Figure 1 | Collective excitations of a material containing a condensate of excitons.** (**A**) Excitons spontaneously condense when electrons and holes bind between two bands whose extrema lie near the chemical potential, separated by wave vector $q_0$,. (**B**) Exciton condensation leads to a modulation in the charge density with period $2\pi/q_0$. (**C**) Sketch of the collective excitations of an exciton condensate in a solid with lattice degrees of freedom. Both electronic and lattice modes soften at $T_C$. At $q_0$ hybrid modes are formed, only one of them reaching zero energy. (**D**) Sketch of the excitations of a conventional Peierls charge density wave, in which the only soft mode is a phonon.

**Figure 2 | Consistency between M-EELS data and previous studies of TiSe$_2$.** (**A**) Elastic ($\omega=0$) momentum map of TiSe$_2$ at room temperature showing the Brillouin zone boundaries (light brown lines) and the (1,0) Bragg peak. (**B**) Same map as panel (A) showing the appearance, at $T < T_C$, of superlattice reflections at (0.5,0) and (0.5,0.5), consistent with previous x-ray and neutron scattering studies (*17,23,36*). These reflections signify the development of a macroscopic population of excitons. (**C**) M-EELS spectra at $q = 0$ showing an electronic mode at 83 meV that shifts and sharpens below $T_C$ (spectra have been offset vertically for clarity), in quantitative agreement with previous studies using infrared spectroscopy (*31,32,33*).

**Figure 3 | Momentum dependence of the valence plasmon in TiSe$_2$ for different temperatures.** (**A**) Normal state M-EELS spectra showing a plasmon at $\omega = 83$ meV that exhibits conventional, Lindhard-like dispersion (*30*). (**B**) Spectra near $T_C$ where electronic and lattice excitations cease to be resolvable. The electronic mode at this temperature reverses its dispersion, going soft at $q = (0.5, 0)$. (**C**) Spectra at $T = 100$ K, showing hardening of the electronic mode. (**D**) $T = 17$ K, showing a fully hardened, nondispersive electronic mode at $\omega = 47$ meV.

**Figure 4 | Summary of the momentum dependence of the soft plasmon mode in TiSe$_2$.** Dispersion curves along the (1, 0) momentum direction were determined by fitting the raw spectra in Fig. 3 (*35*). The error bars represent statistical and systematic contributions, the latter determined by applying several different fit models to the data (*35*). Thick, vertical bars denote spectra that exhibit a power law form, $S \sim \omega^{-1}$, instead of a discernable peak. For comparison, the solid lines show the phonon dispersions along the $A \rightarrow L$ direction at *T* = *T$_C$*, reproduced from Ref. (*41*). Only the *L$_2$* mode (black line) participates in the phase transition. The plasmon behavior is that of the soft mode of a phase transition (see Fig. 1C), demonstrating the condensation of electron-hole pairs at *T$_C$*.

**Acknowledgements.**

We thank C. M. Varma, A. H. MacDonald, G. Baym, and A. J. Leggett for discussions, and I. Bozovic, J. D. Stack, and S. A. Kivelson for feedback on the manuscript. This work was funded by the Gordon and Betty Moore Foundation's EPiQS Initiative through Grant GBMF-4542. Development of the M-EELS instrument was supported by the DOE Center for Emergent Superconductivity, an Energy Frontier Research Center funded by the DOE Office of Science under award no. DE-AC02-98CH10886, with equipment provided by DOE grants DE-FG02-08ER46549 and DE-FG02-07ER46453. J.v.W. acknowledges support from a VIDI grant from the Netherlands Organization for Scientific Research (NWO). G. J. M. and E. F. acknowledge support from DOE grant no. DE-SC0012368. T. C. C. acknowledges support from DOE grant no. DE-FG02-07ER46383. F. F. acknowledges support from a Lindemann Trust Fellowship of the English Speaking Union. The data presented in this manuscript are available from the authors upon reasonable request.

**Supplementary Materials**

www.sciencemag.org/content/???/????/????/suppl/DCI

Materials and Methods

Supplementary Text

Figs. S1-S6

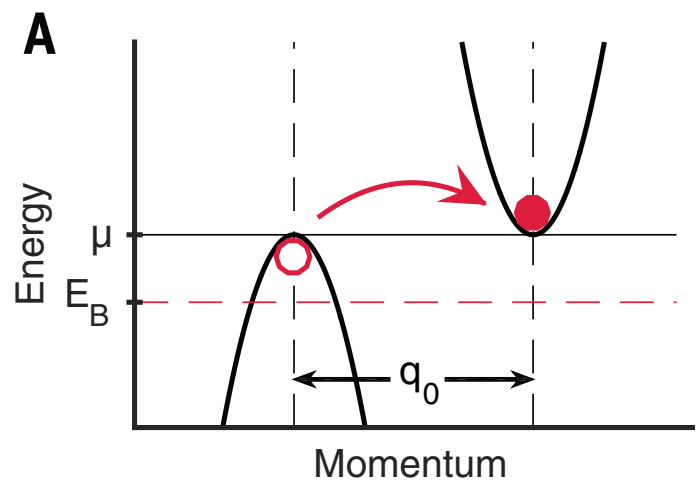
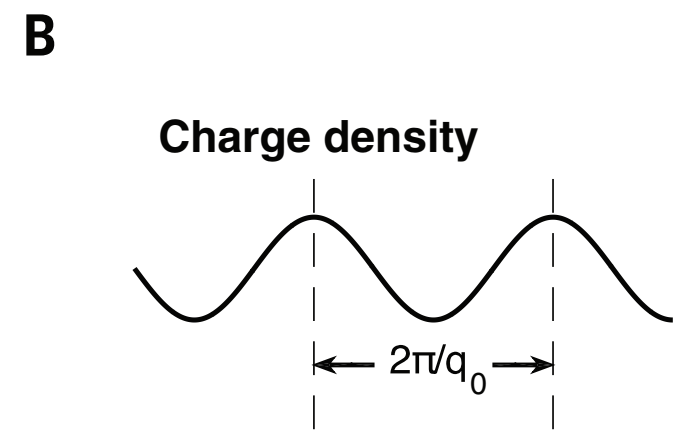
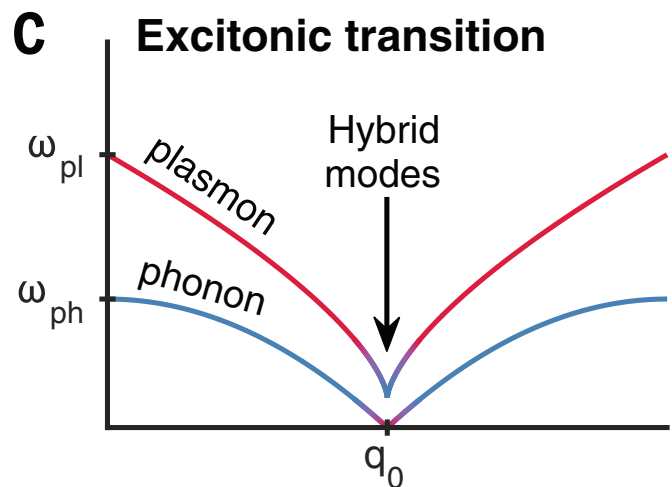
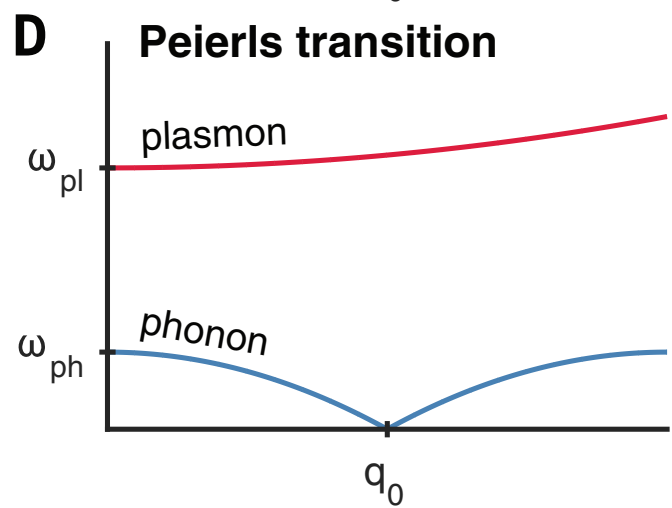

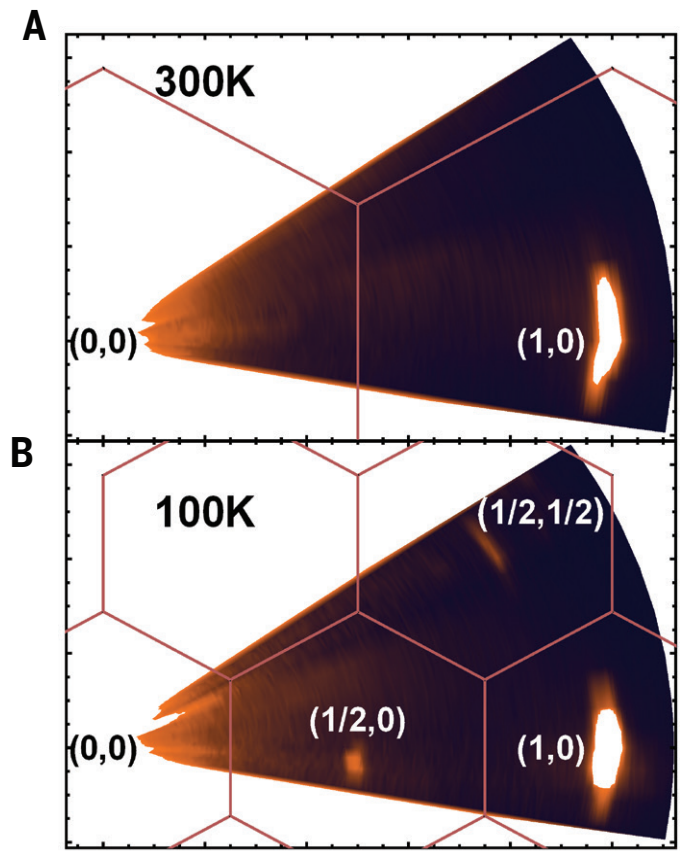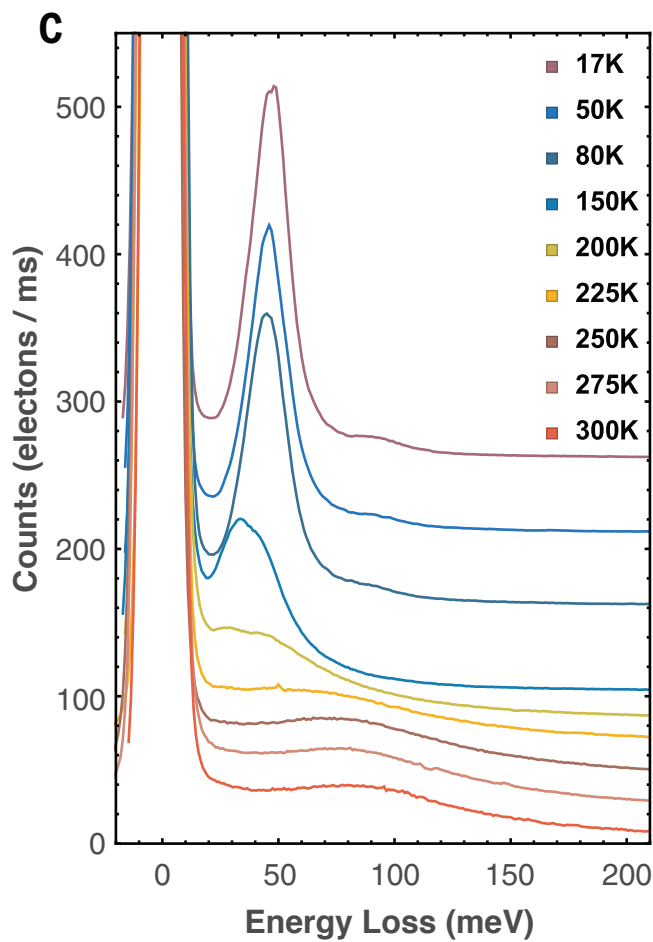

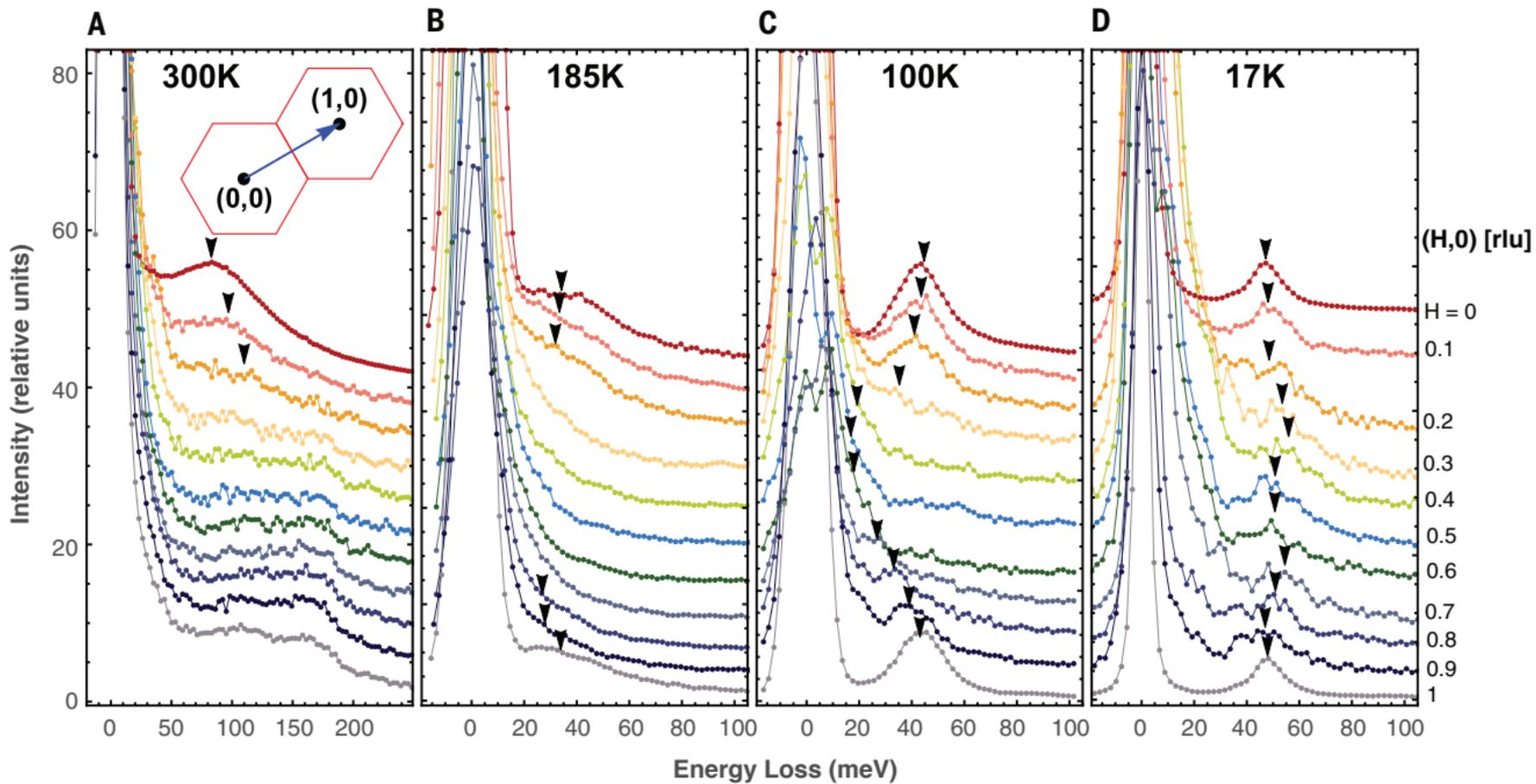

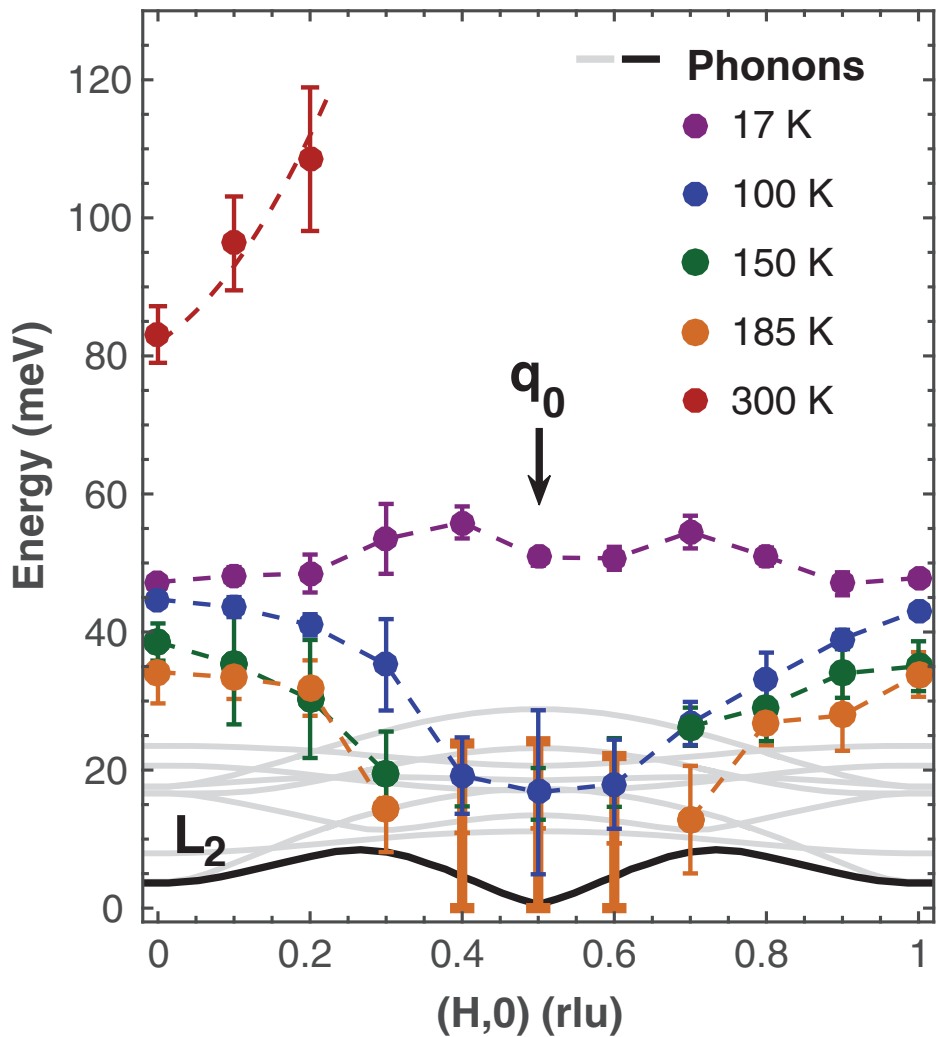

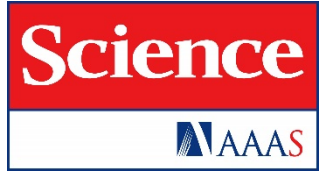

# Supplementary Materials for

## Signatures of exciton condensation in a transition metal dichalcogenide

A. Kogar, M. S. Rak, S. Vig, A. A. Husain, Y. I. Joe, L. Venema, G. J. MacDougall, T. C. Chiang, E. Fradkin, F. Flicker, J. van Wezel, P. Abbamonte

correspondence to: abbamonte@mrl.illinois.edu or kogar2@illinois.edu

**This PDF file includes:**

Materials and Methods
Supplementary Text
Figs. S1 to S6

**Materials and Methods**

Crystal Growth

TiSe$_2$ crystals were grown using the iodine vapor transport method. Ti powder and slight excess Se powder were loaded into a vacuum sealed quartz tube with trace amounts of iodine. The tube was heated to 570-640 degrees Celsius for 6 hours and maintained in a temperature gradient for seven days and cooled to room temperature over a period of 12 hours.

M-EELS Measurements

M-EELS measurements were taken with a conventional HR-EELS spectrometer which was retrofitted with a custom low-temperature seven-axis ultra-high vacuum compatible goniometer. The goniometer was aligned so that the sample rotation axis intersected the spectrometer rotation axis at a single point. Sample orientation was accomplished by identifying two non-collinear Bragg peaks (e.g. (1,0) and (0,1)) to construct an *in situ* orientation matrix translating between diffractometer angles and reciprocal space. Samples were cleaved *in situ* at room temperature and then cooled to the desired temperature. Spectra were obtained at an incident beam energy of 50 eV with an energy resolution between 4-6 meV and a momentum resolution of 0.03 inverse Angstroms. Electron energy-loss scans were accomplished by moving the sample rotation and detector positions to keep the out of plane momentum transfer fixed while simultaneously scanning the voltage on the analyzer. The out-of-plane momentum transfer was fixed to 3.75 reciprocal lattice units to enable the spectrometer to reach the in-plane Brillouin zone boundary at 50 eV beam energy while maintaining an incident beam angle of at least 30 degrees with respect to the sample surface. Further instrumentation-related specifics can be found in Ref. *(18)*.

**Supplementary Text**

Phonon features in the M-EELS spectra of TiSe$_2$

In addition to the electronic mode, several phonon features are visible in the M-EELS spectra. The most prominent, summarized in Fig. S1, is a transverse acoustic (TA) phonon that (as expected) resides at zero energy at $q = (0, 0)$ and $q = (1, 0)$, and disperses to a maximum energy of 11 meV at $q = (0.5, 0)$. This phonon involves displacement of charged ions perpendicular to the surface, so has a large M-EELS cross section and is particularly pronounced in the spectra. The same mode was observed in inelastic neutron studies in 1978 (*34*), which reported a dispersion relation that is quantitatively consistent with our M-EELS measurements (Fig. S1).

In addition to the TA phonon, several weaker phonon modes were visible, which in some spectra appear as a reasonably well-defined peak at 20 meV [see, for example, the $q = (0.9, 0)$ spectrum at 100 K in Fig. 3c of the main manuscript]. In other spectra, these excitations form a featureless continuum spanning the range 10-25 meV. These features are likely combinations of multiple phonon branches, which have been studied extensively (*27,33,34,41*).

More complete data set on the plasmon-like collective mode in TiSe$_2$

The data set in Fig. 3 of the main manuscript was abbreviated for the sake of simplicity. A more complete data set is given in Fig. S2.

Consistency between *q*=0 M-EELS spectra and infrared optics
The manuscript states that M-EELS measurements at $q = 0$ are consistent with measurements of the inverse dielectric function using infrared reflectivity. A side-by-side comparison is presented in Fig. S4.

Determination of dispersion curves and error bars in Figure 4
This section details the method by which the energy points and error bars in Fig. 4 of the main manuscript were obtained. This procedure was applied at each momentum value and temperature.

The main sources of systematic error in this measurement are (1) shift of the elastic line during the experiment due to voltage drifts, (2) ambiguity in how to distinguish the energy of the electronic mode from the TA phonon, particularly in the vicinity of the phase transition, and (3) ambiguity in how to distinguish the electronic mode from the background (phonon) continuum, which is strongest in the 5-20 meV range.

In order to estimate these errors, each spectrum was fit using the least squares method according to two different schemes. Both use a Gaussian profile to fit the elastic peak, a Lorentzian profile to fit each of the inelastic features, and a constant background. The first, scheme A, assumes the presence of two phonon modes, an acoustic phonon—with both an anti-Stokes and a Stokes component—, an optical phonon and the presence of the electronic mode. The second, scheme B, assumes only the presence of the elastic peak and the electronic mode, modeling the phonon background as a single Lorentzian. These two schemes are illustrated in Fig. S5, which shows fits at $q = (0.9, 0)$ and $T = 100$ K. In each of the two fits, the electronic mode energy was defined as the difference between its fit value and the energy of the elastic line,

$$E_{EM}^A = E_{fit}^A - E_0^A \qquad \text{(S1)}$$
$$E_{EM}^B = E_{fit}^B - E_0^B$$

where $E_{fit}$, and $E_0$ are the fit energies of the electronic mode and elastic energy to the raw data in each of schemes A and B. The mode energy was taken to be the average of these two values, i.e.,

$$E_{EM} = \frac{E_{EM}^A + E_{EM}^B}{2}. \qquad \text{(S2)}$$

Each of the fit values in Eq. S1 exhibits a statistical source of error, which is defined by the corresponding diagonal values of the variance-covariance matrix generated in the fit. The total statistical error for each of the two schemes was determined by adding the two in quadrature,

$$\delta_A = \sqrt{\left(\delta_{EM}^A\right)^2 + \left(\delta_0^A\right)^2} \tag{S3}$$

$$\delta_B = \sqrt{\left(\delta_{EM}^B\right)^2 + \left(\delta_0^B\right)^2}$$

where $\delta_{EM}$ and $\delta_0$ represent the errors in the mode energy and elastic energy, respectively, in each of the two schemes.

Finally, the systematic error was estimated by taking the deviation from the average value of the electronic mode obtained from the two fitting schemes:

$$\sigma_A = \left|E_{EM}^A - E_{EM}\right| \tag{S4}$$

$$\sigma_B = \left|E_{EM}^B - E_{EM}\right|.$$

The final error bars in Figure 4 of the main manuscript were determined by adding both the statistical and systematic errors in quadrature,

$$\sigma = \sqrt{\delta_A^2 + \delta_B^2 + \sigma_A^2 + \sigma_B^2}. \tag{S5}$$

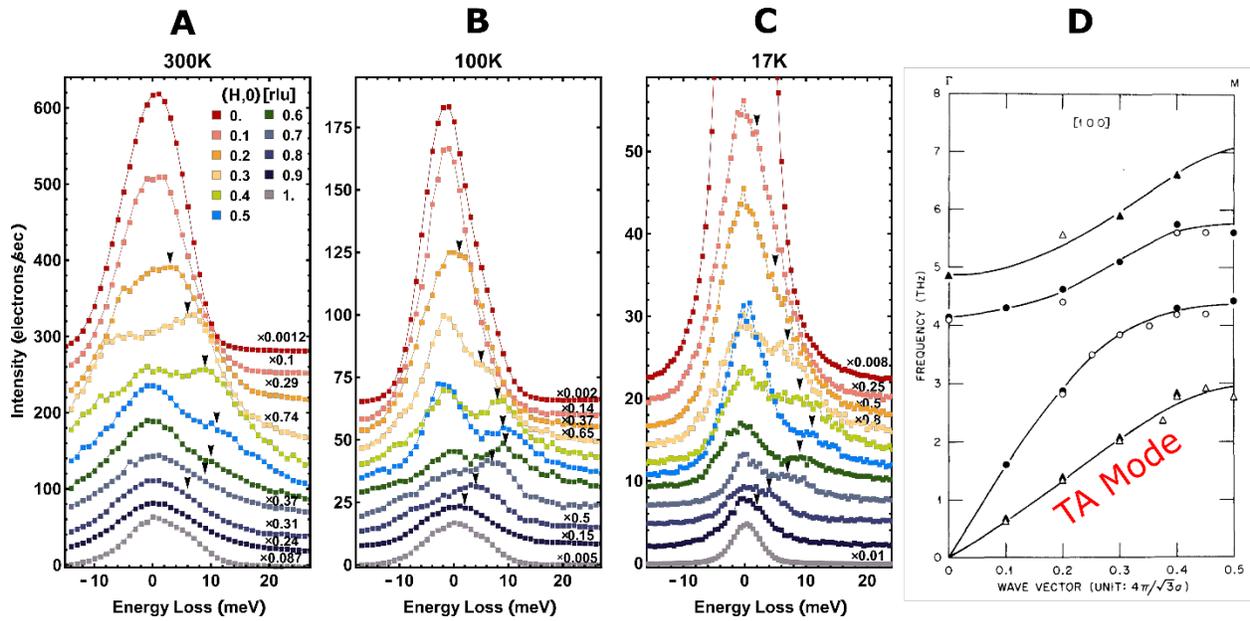

**Fig. S1. Momentum dependence of the M-EELS spectra in the phonon region, taken on the momentum interval (0,0) → (1, 1).** (**A-C**)**,** Individual spectra for 300 K, 100 K, and 17 K, showing the dispersion of a TA phonon in TiSe$_2$. Individual spectra have been scaled by the indicated factors and vertically offset for clarity. This phonon does not participate in the CDW and exhibits no apparent change through $T_C$. Note that the elastic ($\omega = 0$) feature at $H = 0.5$ is enhanced below $T_C$, indicating development of the excitonic order parameter. (**D**) Phonon dispersion curves from inelastic neutron scattering, reproduced from Ref. (*34*). The dispersion of the TA mode is quantitatively consistent with M-EELS studies, showing consistency between the two techniques.

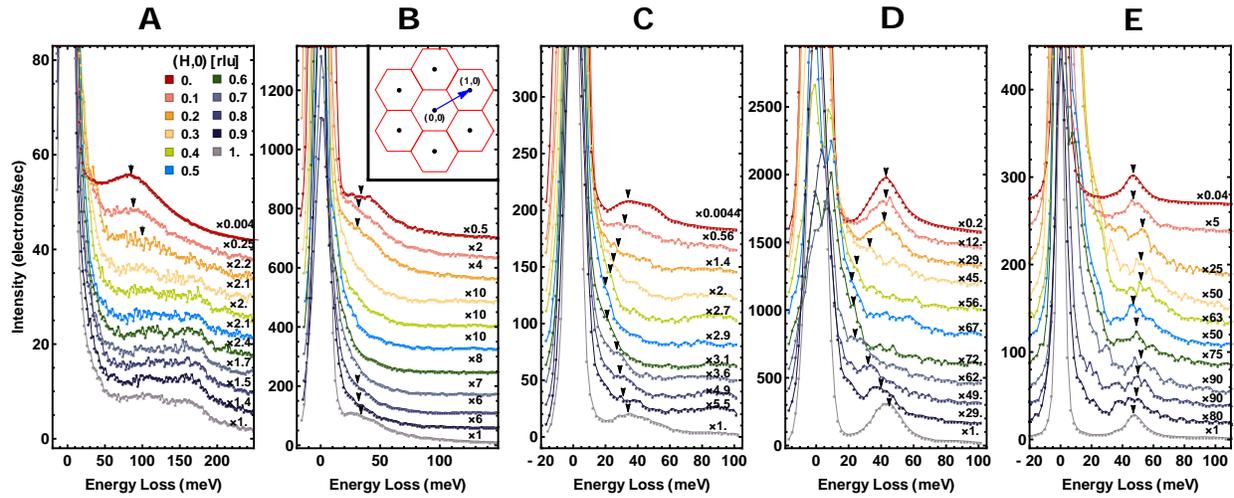

**Fig. S2. Complete M-EELS data set on the (0,0) → (1,0) interval, including the spectra at 150 K and the intensity scale factors** (**A**) Normal state M-EELS spectra showing the valence plasmon at ω = 83 meV that exhibits conventional, Lindhard-like dispersion. The number next to each spectrum represents a factor used to scale the intensity so all the data could be showed on the same scale. (**B**) Spectra near $T_C$ where electronic and lattice excitations cease to be resolvable. The electronic mode at this temperature reverses its dispersion, going soft at q = (0.5, 0). (**C**) Spectra at T = 150 K, showing slight hardening of the plasmon. (**D**) T = 100 K. (**E**) T = 17K, showing a fully hardened, weakly dispersive electronic mode at ω = 47 meV.

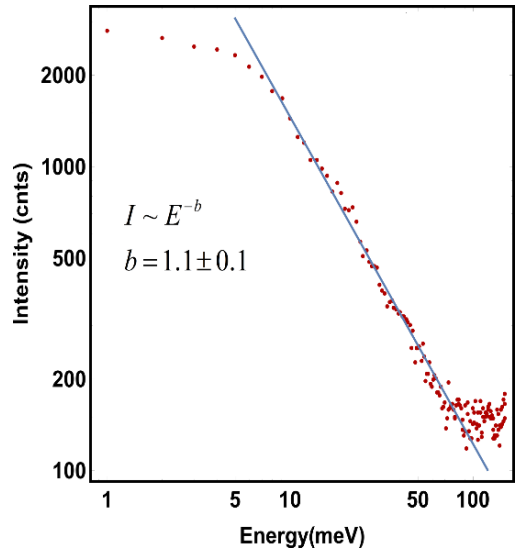

**Fig. S3. M-EELS spectrum at $q = (0.5, 0)$ and $T = T_C$ plotted on a log-log plot**. The purpose of this plot is to illustrate that that the spectrum exhibits a power law form. We state in the manuscript that, at $T \sim T_C$, the electronic mode becomes gapless. What this means, quantitatively, is that there is no visible energy scale in the problem, i.e., to within the limit of the resolution the M-EELS spectrum just exhibits a power law in energy. As shown above, a best fit to this power law gives $S(\omega) \sim \omega^{-1}$. This means that, near $T_C$, the system exhibits the expected dynamical critical fluctuations characteristic of a second order phase transition.

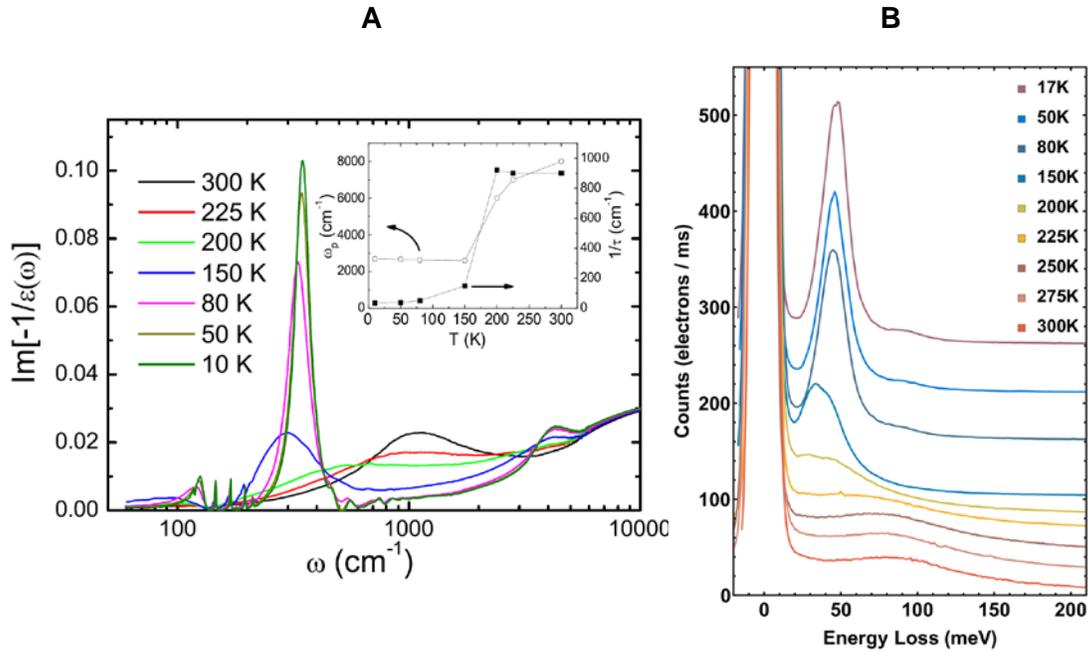

**Fig. S4. Spectrum of collective excitations at *q*=0 as measured with both IR spectroscopy and M-EELS.** (**A**) Frequency-dependent dielectric loss function, −Im[1/ε(*q*,ω)], determined at *q* = 0 from infrared reflectivity measurements of TiSe$_2$ (*33*) (**B**) M-EELS spectrum at *q* = 0. The two techniques see the same fundamental, electronic collective mode, with the same changes in energy, width, and spectral weight when cooling through $T_C$. The agreement is excellent considering that IR is a bulk probe and M-EELS measures only the surface. This measurement establishes that M-EELS measures a quantity closely related to the dielectric loss function.

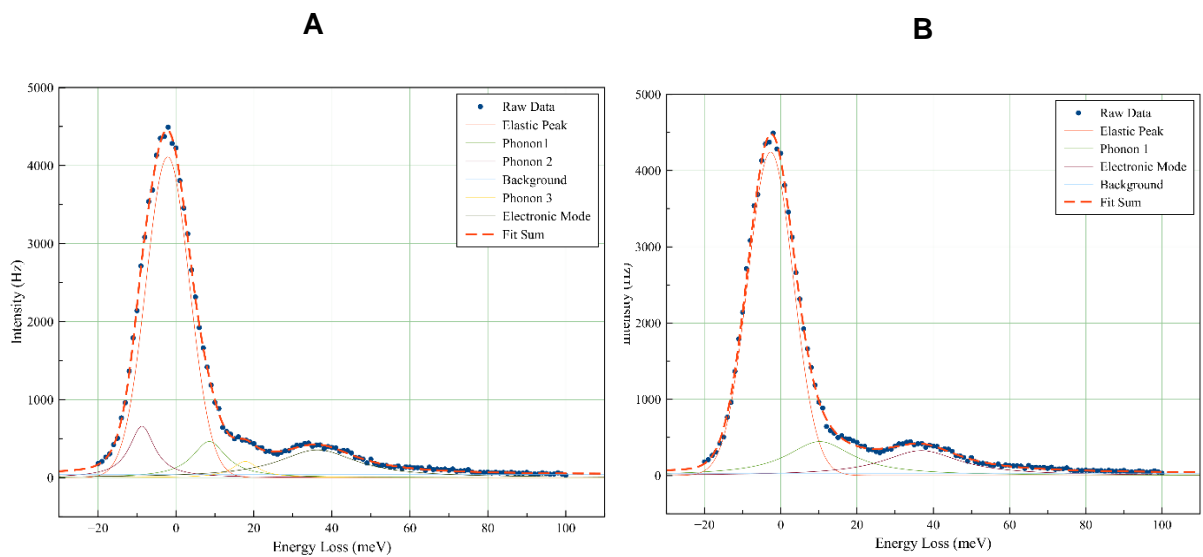

**Fig. S5**. **Example fitting of the M-EELS spectrum at 100 K at q = (0.9,0) using two different schemes.** Both fit schemes use a Gaussian profile to fit the elastic peak, Lorentzian profiles to fit the inelastic features, and a constant background. (**A**) Fit scheme A assumes the presence of three phonon modes, two acoustic—one on the anti-Stokes side and one on the Stokes side—, one optical and the electronic mode. (**B**) Fit scheme B, assumes the presence of only the elastic peak and the electronic mode, while the background phonons are modeled using a single Lorentzian.

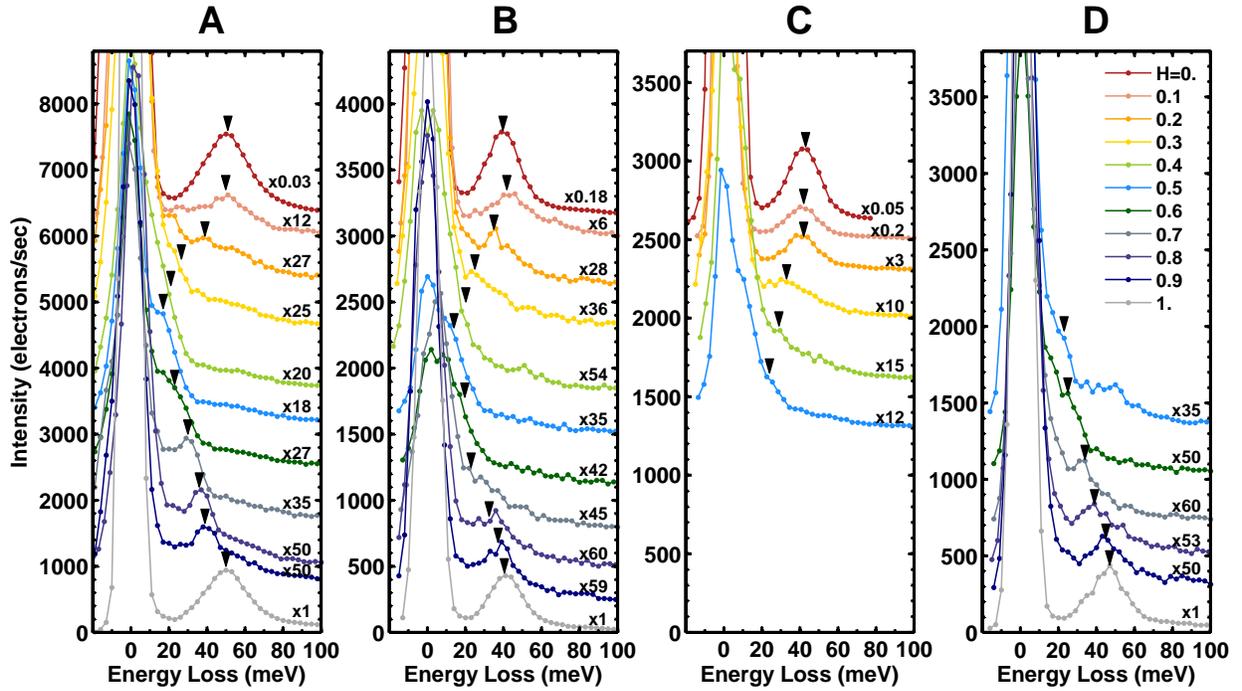

**Fig. S6. Reproducibility of electronic mode dispersion at 100K.** (**A-D**) Electronic collective mode dispersion at 100K on different TiSe$_2$ samples. These spectra were taken on samples different from the one in Fig. 3C of the main manuscript. The dispersion on different samples consistently shows a softening behavior of the electronic mode toward $q_0 = (0.5, 0)$ with a resolvable shoulder at $q_0$ for all spectra taken. Because of slightly different impurity concentrations, the peak energy varies slightly from sample to sample, but the overall behavior is consistent with Fig. 3C in the main manuscript.